\documentclass[showpacs,amsmath,amssymb,twocolumn,prl,superscriptaddress]{revtex4-1}

\usepackage{pifont}
\usepackage{diagbox}
\usepackage{graphicx,epsfig,dsfont,amssymb,amsthm,amsfonts,amsbsy,mathrsfs,amscd}
\usepackage{epsfig}
\usepackage{times}
\usepackage{color}
\usepackage{graphicx}
\usepackage{physics}
\usepackage[noend,noline,ruled,linesnumbered]{algorithm2e}
\usepackage{framed}
\usepackage{tikz}
\usetikzlibrary{calc}
\usetikzlibrary{plotmarks,shapes}
\usepackage{graphicx}
\usepackage{qcircuit,amsmath}
\usepackage{float}
\usepackage{natbib}
\usepackage{booktabs}
\usepackage{pgfplots}
\usepackage[titletoc,title]{appendix}
\usepackage{hyperref}
\usepackage{caption}
\usepackage{subcaption}
\hypersetup{  colorlinks=true, linkcolor=blue, citecolor=red, urlcolor=blue }
\usepackage[singlelinecheck=off,justification=raggedright]{caption}

\newtheorem{lemma}{Lemma}

\newtheorem{theorem}{Theorem}

\usepackage[utf8]{inputenc}

\newcommand{\pbra}[1]{\left( #1 \right)}
\newcommand{\CNF}{{\rm{CNF}}}

\newcommand{\Ocal}{\mathcal{O}}

\newcommand{\Ccal}{\mathcal{C}}

\begin{document}
\title{
Efficient quantum circuit synthesis for SAT-oracle with limited ancillary qubit
}
\author{Shuai Yang}
\affiliation{Institute of Computing Technology, Chinese Academy of Sciences, 100190 Beijing, China}
\affiliation{University of Chinese Academy of Sciences, 100049 Beijing, China}

\author{Wei Zi}
\affiliation{Institute of Computing Technology, Chinese Academy of Sciences, 100190 Beijing, China}
\affiliation{University of Chinese Academy of Sciences, 100049 Beijing, China}

\author{Bujiao Wu}
\affiliation{Center on Frontiers of Computing Studies, Peking University, Beijing 100871, China}

\author{Cheng Guo}
\affiliation{Institute of Computing Technology, Chinese Academy of Sciences, 100190 Beijing, China}
\affiliation{University of Chinese Academy of Sciences, 100049 Beijing, China}

\author{Jialin Zhang}
\affiliation{Institute of Computing Technology, Chinese Academy of Sciences, 100190 Beijing, China}
\affiliation{University of Chinese Academy of Sciences, 100049 Beijing, China}

\author{Xiaoming Sun\footnote{sunxiaoming@ict.ac.cn}}
\affiliation{Institute of Computing Technology, Chinese Academy of Sciences, 100190 Beijing, China}
\affiliation{University of Chinese Academy of Sciences, 100049 Beijing, China}
\email[]{sunxiaoming@ict.ac.cn}

\begin{abstract} 


{ How to implement quantum oracle 
with limited resources raises concerns these days. We design two ancilla-adjustable and efficient algorithms to synthesize SAT-oracle, the key component in solving SAT problems. The previous work takes $2m\!-\!1$ ancillary qubits and $\tilde{O}(m)$ elementary gates to synthesize an $m$ clauses oracle.
The first algorithm reduces the number of ancillary qubits to $2\sqrt{m}$, with at most an eightfold increase in circuit size. 
The number of ancillary qubits can be further reduced to 3 with a quadratic increase in circuit size. The second algorithm aims to reduce the circuit depth. By leveraging of the second algorithm, the circuit depth can be reduced to $\tilde{O}(\log m)$ with $m$ ancillary qubits. 
}

\end{abstract}



\maketitle
\section{Introduction}


Quantum computation has been extensively studied since Feynman first proposed 
in the 1980s \cite{feynman1982simulating}. 
Several quantum algorithms have been proposed which are superior to the best classical algorithms, such as Shor's  algorithm and Grover's algorithm~\cite{Shor1994Polynominal,grover1996fast}. As a result, more and more attention is paid to quantum computation \cite{beigi2020quantum,magniez2007quantum}.
%


In these quantum algorithms, quantum oracles are used to evaluate the value of Boolean function~\cite{nielsen2002quantum}. Here a Boolean function is a function  $f:\{0,1\}^n\to\{0,1\}$.
The function of a quantum oracle is transforming $\ket{x}\ket{c}$ into $\ket{x}\ket{c\oplus f(x)}$~ \cite{nielsen2002quantum}.
To implement these quantum algorithms on quantum devices, we have to decompose the oracle into elementary gates.
Since we are in a noisy intermediate-scale quantum (NISQ) era, the number of qubits and the fidelity and decoherence time of the elementary gate is still at a low level by far \cite{preskill2018quantum,arute2019quantum}.
Although the above quantum oracle can be implemented theoretically, the huge number of quantum resources is unavailable in the NISQ era.
Therefore, it is essential to implement a quantum oracle with as few quantum costs as possible. 


There are several works for the synthesis of quantum oracle \cite{shende2003synthesis,miller2003transformation,wille2009bdd,fazel2007esop}. Those algorithms focus on different representations for Boolean functions. However, for Conjunction Normal Form (CNF) Boolean function, those algorithms need exponential running time to synthesize such an oracle. Here a CNF Boolean function is an AND of several clauses. Each clause is an OR of variables or their negations. 
We denote the quantum oracle of the CNF formula as the SAT-oracle.

The well-known NP-hard problem --- satisfiability (SAT) problem determines whether a CNF is satisfiable  \cite{CNFintro,cook1971complexity,levin1973universal}. 
SAT problems appear in several practical application domains, { such as gene regulatory networks, model checking, electronic design automation, etc \cite{corblin2007sat,mcmillan2003interpolation,kunz1997reasoning}. }
In classical computation and quantum computation, enormous studies aim to solve the SAT problem \cite{schoning1999probabilistic,paturi2005improved,hansen2019faster,dunjko2018computational,leporati2007three}. 
{Those quantum algorithms use {SAT-oracle} to evaluate the value of the CNF Boolean function.} SAT-oracle can also be used in quantum state preparation~\cite{rosenthal2021query}.

Now, we give the definition of the quantum circuit synthesis problem for SAT-oracle. For a given CNF formula $f$ over $n$ variables $x=(x_1,x_2,\cdots,x_n)$, the task is to construct a quantum circuit $\Ccal$ such that $\Ccal\ket{\psi}\ket{c}= \ket{\psi}\ket{c\oplus f(\psi)}.$ 
For convenience, we denote $n$ variables $m$ clauses $k$-CNF (each clause contains at most $k$ variables) as $\CNF_{n,m}^k$. 
Ancillary qubits are widely used in the quantum circuit synthesis and the optimization of quantum circuits.
An idea in \cite{qiskit,campbell2019applying} is to store the value of clauses in the ancillary qubits and then calculate the AND function with a Toffoli gate~\cite{qiskit_alg}. When the ancillary qubits are limited, this algorithm fails.
Inspired by the construction of multi-controlled Toffoli (MCT) in \cite{nielsen2002quantum}, we design an algorithm to synthesize a general quantum AND (OR) circuit for functions rather than variables to conquer the limitation of ancillary qubits. Based on the general AND circuit, we design the size-oriented algorithm to synthesize ${\rm CNF}_{n,m}^k$. The algorithm costs $\ell$ ancillary qubits and $\tilde{O}(km^{1+o(1)})$ elementary gates. The size of the circuit decreases rapidly with the growth of $\ell$. Particularly when $\ell = m^\epsilon$, the circuit size drops to $O(km/\epsilon)$. Then, we introduce depth-oriented algorithm to reduce the depth of the quantum circuit to $\tilde{O}(\log km)$ with $km$ ancillary qubits, where the size increases by a logarithm factor. When the ancillary qubits is limited, the circuit depth is roughly $\tilde{O}(km^{1+o(1)}/\ell)$. The running time of the two algorithms is both $O(km)$.
The experimental results show that with a tolerable (a constant ratio) increase in the size of the quantum circuit, the number of ancillary qubits is reduced { from $2m-1$} to $2\sqrt{m}$ using the size-oriented algorithm. The depth-oriented algorithm significantly reduces the circuit depth of the SAT-oracle. We also give a resource estimate of solving a meaningful SAT problem using Grover's algorithm and our synthesis algorithm.
 

\section{Result and method}


\label{sec:dirty_anci}

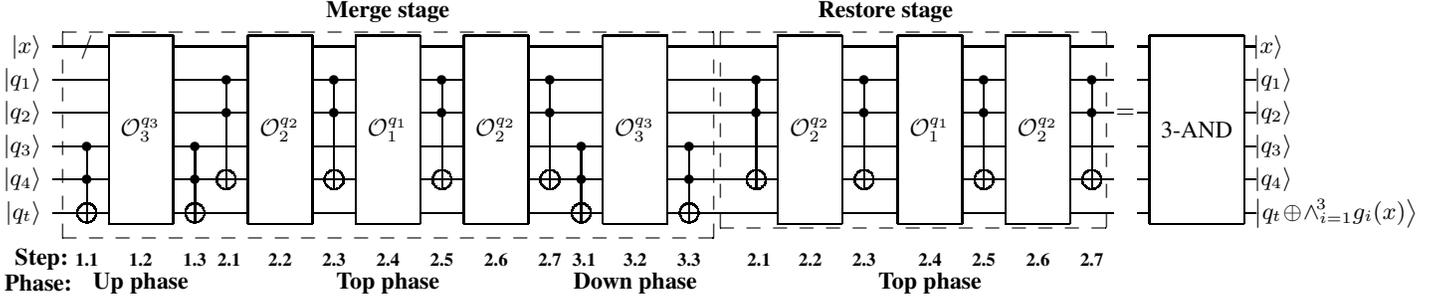
\begin{figure*}[ht]
\Qcircuit @C=0.5em @R=.5em{
\lstick{}&&&&&&&&\textbf{Merge stage}&&&&&&&&&&&&&\textbf{Restore stage}\\
\\
\lstick{\ket{x}} & \qw &\qw{/} &\multigate{5}{\mathcal{O}_3^{q_3}}&\qw
&\qw&\multigate{5}{\mathcal{O}_2^{q_2}}&\qw&\multigate{5}{\mathcal{O}_1^{q_1}}&\qw&\multigate{5}{\mathcal{O}_2^{q_2}}&\qw
&\qw&\multigate{5}{\mathcal{O}_3^{q_3}}&\qw&\qw&\qw \gategroup{3}{3}{8}{15}{1.2em}{--}  %
&\qw&\qw&\multigate{5}{\mathcal{O}_2^{q_2}}&\qw&\qw&\multigate{5}{\mathcal{O}_1^{q_1}}&\qw&\multigate{5}{\mathcal{O}_2^{q_2}}&\qw&\qw&
&&\multigate{5}{\mbox{3-AND}}&\qw&\ket{x}  \gategroup{3}{18}{8}{26}{1.2em}{--} \\
\lstick{\ket{q_1}} &\qw&\qw &\ghost{\mathcal{O}_3^{q_3}}&\qw
&\ctrl{1}&\ghost{\mathcal{O}_2^{q_2}}&\ctrl{1}&\ghost{\mathcal{O}_1^{q_1}}&\ctrl{1}&\ghost{\mathcal{O}_2^{q_2}}&\ctrl{1}
&\qw &\ghost{\mathcal{O}_3^{q_3}} &\qw&\qw&\qw%
&\qw&\ctrl{1}&\ghost{\mathcal{O}_2^{q_2}}&\ctrl{1}&\qw&\ghost{\mathcal{O}_1^{q_1}}&\ctrl{1}&\ghost{\mathcal{O}_2^{q_2}}&\ctrl{1}&\qw&
&&\ghost{\mbox{3-AND}}&\qw&~\ket{q_1}\\
\lstick{\ket{q_2}} &\qw&\qw&\ghost{\mathcal{O}_3^{q_3}}&\qw
&\ctrl{2}&\ghost{\mathcal{O}_2^{q_2}}&\ctrl{2}&\ghost{\mathcal{O}_1^{q_1}}&\ctrl{2}&\ghost{\mathcal{O}_2^{q_2}}&\ctrl{2}
&\qw&\ghost{\mathcal{O}_3^{q_3}} &\qw&\qw&\qw%
&\qw&\ctrl{2}&\ghost{\mathcal{O}_2^{q_2}}&\ctrl{2}&\qw&\ghost{\mathcal{O}_1^{q_1}}&\ctrl{2}&\ghost{\mathcal{O}_2^{q_2}}&\ctrl{2}&\qw&=
&&\ghost{\mbox{3-AND}}&\qw&~\ket{q_2}\\
\lstick{\ket{q_3}} &\qw&\ctrl{1}&\ghost{\mathcal{O}_3^{q_3}}&\ctrl{1}
&\qw&\ghost{\mathcal{O}_2^{q_2}}&\qw&\ghost{\mathcal{O}_1^{q_1}}&\qw&\ghost{\mathcal{O}_2^{q_2}}&\qw
&\ctrl{1}&\ghost{\mathcal{O}_3^{q_3}}&\ctrl{1}&\qw&\qw%
&\qw&\qw&\ghost{\mathcal{O}_2^{q_2}}&\qw&\qw&\ghost{\mathcal{O}_1^{q_1}}&\qw&\ghost{\mathcal{O}_2^{q_2}}&\qw&\qw&
&&\ghost{\mbox{3-AND}}&\qw&~\ket{q_3}\\
\lstick{\ket{q_4}} &\qw &\ctrl{1}&\ghost{\mathcal{O}_3^{q_3}}&\ctrl{1}
&\targ&\ghost{\mathcal{O}_2^{q_2}}&\targ&\ghost{\mathcal{O}_1^{q_1}}&\targ&\ghost{\mathcal{O}_2^{q_2}}&\targ
&\ctrl{1}&\ghost{\mathcal{O}_3^{q_3}}&\ctrl{1}&\qw&\qw%
&\qw&\targ&\ghost{\mathcal{O}_2^{q_2}}&\targ&\qw&\ghost{\mathcal{O}_1^{q_1}}&\targ&\ghost{\mathcal{O}_2^{q_2}}&\targ&\qw&
&&\ghost{\mbox{3-AND}}&\qw&~\ket{q_4}\\
\lstick{\ket{q_t}} &\qw &\targ&\ghost{\mathcal{O}_3^{q_3}}&\targ
&\qw&\ghost{\mathcal{O}_2^{q_2}}&\qw&\ghost{\mathcal{O}_1^{q_1}}&\qw&\ghost{\mathcal{O}_2^{q_2}}&\qw
&\targ&\ghost{\mathcal{O}_3^{q_3}}&\targ&\qw&\qw%
&\qw&\qw&\ghost{\mathcal{O}_2^{q_2}}&\qw&\qw&\ghost{\mathcal{O}_1^{q_1}}&\qw&\ghost{\mathcal{O}_2^{q_2}}&\qw&\qw&
&&\ghost{\mbox{3-AND}}&\qw&~~~~~~~~~~~~~~~~~~~~~~\ket{q_t\! \oplus \! \wedge_{i=1}^{\!3} g_i(x)}\\
\\
\\
\textbf{Step:}~~~~&&\textbf{\scriptsize1.1}&\textbf{\scriptsize1.2}&\textbf{\scriptsize1.3}&~\textbf{\scriptsize2.1}&\textbf{\scriptsize2.2}&\textbf{\scriptsize2.3}&\textbf{\scriptsize2.4}&\textbf{\scriptsize2.5}&\textbf{\scriptsize2.6}&\textbf{\scriptsize2.7}&~\textbf{\scriptsize3.1}&\textbf{\scriptsize3.2}&\textbf{\scriptsize3.3}&&&&~\textbf{\scriptsize2.1}&\textbf{\scriptsize2.2}&\textbf{\scriptsize2.3}&&\textbf{\scriptsize2.4}&\textbf{\scriptsize2.5}&\textbf{\scriptsize2.6}&\textbf{\scriptsize2.7}\\
\\
\textbf{Phase:}~~~~~&&&\textbf{Up phase}&&&&&\textbf{Top phase}&&&&&\textbf{Down phase}&&&&&&&&&\textbf{Top phase}\\
}
\caption{Construction of $3$-GAND circuit. $\ket{x}$ is input qubits, $\ket{q_t}$ is target qubit and $\ket{q_{1}},\cdots,\ket{q_4}$ are ancillary qubits.  Here $\mathcal{O}_i^{q}$ is the $\mathcal{O}_i$ on the target qubit $q$.}
\label{fig:mer_cir}
\end{figure*}
Consider a general AND problem: 
given $p$ Boolean functions $g_1(x),g_2(x),\cdots,g_p(x)$ and the corresponding quantum oracles $\Ocal_i\ket{x}\ket{c}=\ket{x}\ket{c\oplus g_i(x)}$. The goal is to construct a quantum circuit $p\mbox{-GAND}$ such that $p\mbox{-GAND}\ket{x}_n\ket{q}_\ell \ket{c}=\ket{x}_n\ket{q}_\ell \ket{c\oplus(\bigwedge_{i=1}^p g_i(x))}$ with $\ell$ ancillary qubits. We call this general $p$-AND function and denote $p\mbox{-GAND}$ as the general $p$-AND circuit. We can define the general $p$-OR circuit similarly.
\begin{lemma}[$p$-GAND circuit]

For any natural number $p$, general $p$-AND circuit can be implemented with $2p$ {dirty} ancillary qubits, $O(p)$ Toffoli gate and 4 calls of each $\Ocal_i$.

\label{lem:dirty}
\end{lemma}
Due to the space constraints, we briefly introduce the structure of the circuit implemented by the algorithm~\cite{supplement}. We give an example as illustrated in Figure~\ref{fig:mer_cir}.
The construction of $p\mbox{-GAND}$ is divided into two stages: the merge stage and the restore stage. The merge stage contains $2p-3$ steps, and the restore stage is to repeat steps $2$ to $2p-4$ of the merge stage. Those $2p-3$ steps can be further divided into three phases: \emph{Up phase}, \emph{Top phase}, and \emph{Down phase}. 

\begin{enumerate}
    \item \emph{Up phase}: In this phase, we add the information in the ancillary qubits to corresponding qubits, which can help to eliminate unexpected information. {Up phase contains the first $p-2$ step.} Step $i$ contains 3 sub-steps: 2 Toffoli gates and a call of $\Ocal_i$. 
    \item \emph{Top phase}: In this phase, we implement a circuit that merges two clauses and stores the result in an ancillary qubit. Top phase only contains step $(p\!-\!1)$, which has seven sub-steps (4 Toffoli gates, 2 calls of $\Ocal_2$ and 1 call of $\Ocal_1$). 
    
    \item \emph{Down phase}: In this phase, we merge all the clauses in the ancillary qubits. With the help of \emph{Up phase}, the target qubit stores the value of input CNF. There are 3 sub-steps in each step $i\in\{p,p+1,\cdots,2p-3\}$, which are 2 Toffoli gates and a call of oracle $\Ocal_{2p-i}$.
\end{enumerate}







By recursively calling this circuit, {we design two efficient algorithms to implement SAT-oracles.} 
The input CNF will be divided into several blocks recursively. In the innermost sub-block, which is numbered $1$st level block, we use the general $k$-OR circuit to generate a sub-CNF. The circuit to construct $i$-th level block is used as an oracle in $(i+1)$-th level block.
{We give the pseudo-code of size-oriented algorithm and depth-oriented algorithm in the algorithm \ref{alg:siz} and \ref{alg:dep} respectively.}

\SetNlSty{textbf}{}{\quad}
\begin{algorithm}[ht]
  \SetStartEndCondition{ }{}{}%
\SetKwProg{Fn}{def}{\string:}{}
\SetKwFunction{Range}{range}
\SetKw{KwTo}{in}\SetKwFor{For}{for}{\string:}{}%
\SetKwIF{If}{ElseIf}{Else}{if}{:}{elif}{else:}{}%
\SetKwFor{While}{while}{:}{fintq}%
\AlgoDontDisplayBlockMarkers\SetAlgoNoEnd\SetAlgoNoLine%
  
  \SetKwFunction{Clause}{\bf Clause}
  \SetKwFunction{MCT}{\bf MCT}
  \SetKwProg{Fn}{}{:}{}
  \SetKwInOut{Input}{input}
  \SetKwInOut{Output}{output}
  \Input{A ${\rm CNF}_{n,m}^k$ instance $f = \bigwedge_{a=1}^m C_a$ and the number of ancillary qubits $\ell\ge 3$.}
  \Output{A circuit $\Ccal$ such that $\Ccal|x\rangle \ket{0}_{\ell} |c\rangle = |x\rangle \ket{0}_{\ell} |c\oplus f(x)\rangle,\forall x \in \{0,1\}^{n}.$}
  \BlankLine
  {$d\gets\log_{\ell/2} m,s\gets \frac{2m}{\ell}$}\; 
  
  \For{$j$ in 1 to $\ell/2$}{
     
     {\qquad\Clause($(j\!-\!1)s \!+\! 1, js,$ Ancilla$[j],~d\!-\!1$)}\;
  }
  \MCT(Ancilla, Target)\;
  
    \For{$j$ in 1 to $\ell/2$}{
     
     {\qquad\Clause($(j\!-\!1)s \!+\! 1, js,$ Ancilla$[j],~d\!-\!1$)}\;
     
    \BlankLine
    \BlankLine
    \BlankLine
      \Fn{\Clause{SId,EId,Target,Depth}}{
  	\If{(Depth=0)}{
  	 \qquad Synthesize clauses on target qubit\;
	\qquad {\KwRet;}\\
    }
	{$s \gets ($EId$-$SId$)/(\ell/2+1)$\;}
	{Apply $(2\ell\!+\!1)$-GAND circuit, where $O_i$ is synthesized by \Clause(SId+$(i\!-\!1)s$, SId$\!+\!is\!-\!1$, Ancilla[$i$], Depth$-1$)\;}
	{\KwRet;}
  }
  }
\caption{Size-oriented algorithm}\label{alg:siz} 
\end{algorithm}

\SetNlSty{textbf}{}{\quad}
\begin{algorithm}[ht]
  \SetStartEndCondition{ }{}{}%
\SetKwProg{Fn}{def}{\string:}{}
\SetKwFunction{Range}{range}
\SetKw{KwTo}{in}\SetKwFor{For}{for}{\string:}{}%
\SetKwIF{If}{ElseIf}{Else}{if}{:}{elif}{else:}{}%
\SetKwFor{While}{while}{:}{fintq}%
\AlgoDontDisplayBlockMarkers\SetAlgoNoEnd\SetAlgoNoLine%
  
  \SetKwFunction{Clause}{\bf Clause}
  \SetKwFunction{MCT}{\bf MCT}
  \SetKwProg{Fn}{}{:}{}
  \SetKwInOut{Input}{input}
  \SetKwInOut{Output}{output}
  \Input{A ${\rm CNF}_{n,m}^k$ instance $f = \bigwedge_{a=1}^m C_a$ and the number of ancillary qubits $\ell\ge 3$.}
  \Output{A circuit $\Ccal$ such that $\Ccal|x\rangle \ket{0}_{\ell} |c\rangle = |x\rangle \ket{0}_{\ell} |c\oplus f(x)\rangle,\forall x \in \{0,1\}^{n}.$}
  \BlankLine
  {$S\gets\max\{k/\log \ell,1\},d\gets\log_{\ell/2(S+1)}m, s\gets \frac{2(S+1)m}{\ell}$}\;
  {Divide the ancillary qubits into 3 parts $q_{mem}, q_{dirty}, q_{clean}$, size of each part is $(S-1)\ell / (S+1)$, $\ell/(S+1)$, $\ell/(S+1)$}\;
  
  \For{$j$ in 1 to $\frac{\ell}{2(S+1)}$}{
     
     {\qquad\Clause($(j\!-\!1)s + 1, js, q_{dirty}[j],d-1$)}\;
  }
  \MCT($q_{dirty}$,Target)\;
  
    \For{$j$ in 1 to $\frac{\ell}{2(S+1)}$}{
     
     {\qquad\Clause($(j\!-\!1)s + 1, js, q_{dirty}[j],d-1$)}\;
     
    \BlankLine
    \BlankLine
    \BlankLine
      \Fn{\Clause{SId,EId,Target,Depth}}{
  	\If{(Depth=0)}{
\qquad{Copy the input to $q_{mem}$;\\}
\qquad{Use the input and $q_{mem}$ to synthesize  $\ell/(S+1)$ clauses in parallel at $q_{dirty}$;\\}
\qquad{Use the $q_{clean}$ to merge all the clauses in parallel;\\}
\qquad{Reset the ancillary qubits}\;

	\qquad {\KwRet;}\\
    }
	{$s \gets \frac{2(S+1)(EId-SId)}{\ell}$\;}
	{Apply the GAND circuit, where $O_i$ is synthesized by \Clause(SId+$(i\!-\!1)s$, SId$+is\!-\!1,~q_{dirty}[i]$, Depth$-1$)\;}
	{\KwRet;}
  }
  }
\caption{Depth-oriented algorithm}\label{alg:dep} 
\end{algorithm}

  

\begin{theorem}
By applying algorithm \ref{alg:siz}, 
any instance of $\CNF_{n,m}^k$ can be implemented by an $O(n(km/n)^{1+\log_{\ell/2+1} 4})$-size circuit with $\ell$ ancillary qubits. 
\label{thm:siz}
\end{theorem}
The {size-oriented} algorithm performs well in the size of the circuit. 
When the ancillary qubits are clean (the initial state is $\ket{0}$), we design depth-oriented algorithm to further reduce the circuit depth, with a little increase in the circuit size. { We partition the ancillary qubits into 3 registers, $q_{mem}$, $q_{dirty}$ and $q_{clean}$. The qubits in $q_{mem}$ and $q_{clean}$ are clean at the beginning. The synthesis framework is implemented with the $q_{dirty}$ register. The difference between the two algorithms is in the innermost recursion, where we use the $q_{mem}$ and $q_{clean}$ to parallel the quantum circuit.} 



\begin{theorem}
{By applying algorithm \ref{alg:dep}, } any instance of $\CNF_{n,m}^k$ can be implemented by  an $O\left(k\left(\frac{mS}{\ell}\right)^{1+c}\log\ell\right)$-depth circuit, where $\ell$ is the number of ancillary qubits, $S=\max\{\frac{k}{\log \ell},1\}$ and $c= \log_{\ell/S} 4$.
\label{thm:depth}
\end{theorem}
The depth of the quantum circuit is declined rapidly with the growth of the number of ancillary qubits. Some numerical experiments are designed to show the performance of our algorithm in the following subsection. We also point out the asymptotically lower bound for synthesizing the CNF formula by counting. 
{
The classical running time is the same as the number of calls to the clause function, which is polynomial to the input size.
Using the counting method, we show the lower bound of the SAT-oracle synthesis problem.}

\begin{theorem}
There exists an instance of ${\rm CNF}_{n,m}^k$, such that any quantum circuits approximating it with error $\varepsilon\! <\! \frac{\sqrt{2}}{2}$ must have size at least $\Omega\pbra{km}$. 
\label{thm:sizeLowerBound}
\end{theorem}


Combining the Theorem~\ref{thm:siz} and Theorem~\ref{thm:sizeLowerBound}, we see that our algorithm is asymptotically optimal when the number of ancillary qubits is $\Omega((km)^\epsilon)$ for any $\epsilon > 0$.

Both two algorithms show significant advantages compare to the previous work. For the size-oriented algorithm, we can reduce the number of ancillary qubits to $O(\sqrt{m})$ with only a constant ratio in size. For the depth-oriented algorithm, we use $O(\log km)$ ancillary qubits to construct the same depth circuit generated by qiskit~\cite{supplement}.


    

We use random $k$-$\CNF$ as the experimental benchmark to test the performance of different algorithms. 
To sample a $\CNF_{n,m}^k$, we first randomly sample $k$ variables from the input variables, then randomly choose the variables or the negations of the variables. After two steps, a clause of $\CNF^k_{n,m}$ is generated. Then, repeating the first two steps $m$ times, we generate a random $\CNF^k_{n,m}$. 
We randomly sample 100 $\CNF$s of different parameters and use the average quantum cost to measure the performance of quantum synthesis algorithms. 
{The width of the $\CNF$ $k$ is 3 and 4, and the number of variables $n$ is 40, 80, 400, and 800.} The number of clauses $m$ we choose in this manuscript is determined by the number of variables $n$ and the width of $\CNF$ $k$. When $k$ is 3 and 4, there are $m=\lfloor 4.267n \rfloor$ and $m=\lfloor 9.931n \rfloor$, respectively, which is called SAT phase transition \cite{mitchell1992hard}. Our algorithm is suitable for all the input. To evaluate the quantum cost to conquer the most difficult SAT instance, we choose such a specific $m$ in our experiments.
Here, the size and the depth of the quantum circuit are considered appropriate quantum costs in the NISQ era.
To verify the relationship between the number of ancillary qubits $\ell$ and the quantum cost of our two algorithms, we choose several different $\ell$. Some are near to the $n$, and others are near to $2m-1$. 
The results of different widths (the number of variables in a clause) seem similar. For convenience, we plot the result of $4$-CNF in the Figure \ref{fig:fig_our_result}, which is appropriate to show the performance of our algorithms. 

\pgfplotsset{
  compat=1.3,
  tick label style={font=\tiny},
  label style={font=\tiny},
  width=6.5cm,
  height=5.5cm,
  every axis/.append style={legend style={ font=\tiny, mark options={scale=0.5} }}}
\begin{figure}[htbp]
    \begin{subfigure}[b]{0.5\textwidth}
    \captionsetup{justification=centering}
	\begin{tikzpicture}[every mark/.append style={mark size=1pt}]
		\begin{axis}[name=plot1,height=4cm,width=3.8cm,ylabel={\#ancillary qubits$(\ell)$},xlabel={size},legend cell align=left,legend pos=outer north east]
			\addplot [blue,mark=diamond*]coordinates{
				( 194528.28 , 40 )
				( 176503.88 , 80 )
				( 162319.4 , 120 )
				( 138216.2 , 160 )
				( 118851.84 , 200 )
				( 84852.2 , 280 )
				( 79973.72 , 320 )
				( 75096.16 , 360 )
				( 70491.32 , 400 )
				( 43644.32 , 793 )
			};
			\addplot [blue,mark=triangle*] coordinates{
				( 391760.16 , 80)
				(  354217.76 , 160)
				(  325390.48 , 240)
				(  276886.2 , 320)
				(  237989.64 , 400)
				(  169815.08 , 560)
				(  160056.88 , 640)
				(  150296.96 , 720)
				(  141176.2 , 800)
				(  87333.2 , 1587)
			};
			\addplot[blue,dashed, every mark/.append style={solid, fill=white},mark=square*] coordinates{
				( 51576.32 , 793 )
			};
			\addplot[blue,dashed, every mark/.append style={solid, fill=red},mark=star] coordinates{
				( 103205.2, 1587 )
			};
			\addplot [blue,dashed,line legend, sharp plot,update limits=false] coordinates { (-200,793) (500000,793) };
			\addplot [blue,dashed,line legend, sharp plot,update limits=false] coordinates { (-200,1587) (500000,1587) };
		\end{axis}
		\begin{axis}[name=plot2,at={($(plot1.east)+(1cm,0cm)$)},anchor=west,height=4cm,width=3.8cm,ylabel={\#ancillary qubits$(\ell)$},xlabel={size},legend cell align=left,legend pos=outer north east]
			\addplot [blue,line legend, sharp plot,update limits=false,mark=diamond*]coordinates{(  1969904.76 , 400 )};
			\addplot [blue,line legend, sharp plot,update limits=false,mark=triangle*]coordinates{(  1969904.76 , 400 )};
			\addplot [blue,dashed,line legend, sharp plot,update limits=false,mark=square*,every mark/.append style={solid, fill=white}]coordinates{(  1969904.76 , 400 )};
			\addplot [blue,dashed,line legend, sharp plot,update limits=false,mark=star]coordinates{(  1969904.76 , 400 )};
			\addplot [mark=square*]coordinates{
				(  1969904.76 , 400 )
				(  1776166.68 , 800 )
				(  1630200.4 , 1200 )
				(  1386513.84 , 1600 )
				(  1191458.44 , 2000 )
				(  849985.28 , 2800 )
				(  801170.8 , 3200 )
				(  752373.52 , 3600 )
				(  706892.2 , 4000 )
				(  436945.2 , 7943 )
			};
			\addplot [mark=*]coordinates{
				(  3942175.56 , 800 )
				(  3553221.16 , 1600 )
				(  3260885.28 , 2400 )
				(  2773254.04 , 3200 )
				(  2383028.4 , 4000 )
				(  1699959.08 , 5600 )
				(  1602345.92 , 6400 )
				(  1504749.24 , 7200 )
				(  1413870.04 , 8000 )
				(  873827.04 , 15887 )
			};
			\addplot[black,densely dotted, every mark/.append style={solid, fill=white},mark=*] coordinates{
				( 516323 , 7943 )
			};
			\addplot[black,densely dotted, every mark/.append style={solid, fill=white},mark=otimes*] coordinates{
				( 1032615, 15887 )
			};
			\addplot [black,densely dotted,line legend, sharp plot,update limits=false] coordinates { (-200000,7943) (18000000,7943) };
			\addplot [black,densely dotted,line legend, sharp plot,update limits=false] coordinates { (-200000,15887) (18000000,15887) };
			\legend{Alg.1: $n=40$,Alg.1: $n=80$,Qiskit: $n=40$,Qiskit: $n=80$,Alg.1: $n=400$,Alg.1: $n=800$,Qiskit: $n=400$,Qiskit: $n=800$}
		\end{axis}
	\end{tikzpicture}
		\centering
		\caption{The number of ancillary qubits needs to reach a given size.}
		\end{subfigure}
		\hfill
		\begin{subfigure}[b]{0.5\textwidth}
		\captionsetup{justification=centering}
	\begin{tikzpicture}[every mark/.append style={mark size=1pt}]
		\begin{axis}[name=plot1,height=4cm,width=3.8cm,ylabel={depth},xlabel={\#ancillary qubits$(\ell)$},legend cell align=left,legend pos=outer north east]
			\addplot [blue,mark=diamond*] coordinates{
				( 80 , 38151.4 )
				( 120 , 31781.8 )
				( 160 , 28253.8 )
				( 200 , 25932.2 )
				( 240 , 24693.0 )
				( 280 , 23626.6 )
				( 320 , 22247.4 )
				( 360 , 22132.2 )
				( 400 , 20786.6 )
				( 440 , 20697.0 )
				( 480 , 20581.8 )
				( 520 , 20268.2 )
				( 560 , 19342.6 )
				( 600 , 19342.6 )
				( 640 , 19303.4 )
				( 680 , 19182.6 )
				( 720 , 19020.2 )
				( 760 , 18843.4 )
				( 793 , 18189.0 )
			};
			\addplot [blue,mark=triangle*] coordinates{

				( 80 , 59228.2 )
				( 160 , 42225.0 )
				( 240 , 35265.0 )
				( 320 , 31248.2 )
				( 400 , 28309.8 )
				( 480 , 26809.8 )
				( 560 , 25594.6 )
				( 640 , 24105.8 )
				( 720 , 23755.4 )
				( 800 , 22328.2 )
				( 880 , 22261.0 )
				( 960 , 22056.2 )
				( 1040 , 21479.4 )
				( 1120 , 20298.6 )
				( 1200 , 20438.6 )
				( 1280 , 20421.8 )
				( 1360 , 20250.6 )
				( 1440 , 20054.6 )
				( 1520 , 19816.2 )
				( 1587 , 19005.0 )
			};
			\addplot[blue,dashed, every mark/.append style={solid, fill=white},mark=square*] coordinates{
				( 793 , 38892 )
			};
			\addplot[blue,dashed, every mark/.append style={solid, fill=red},mark=star] coordinates{
				( 1587 , 77798.0 )
			};
			\addplot [blue,dashed,line legend, sharp plot,update limits=false] coordinates { (-200,38892) (2000,38892) };
			\addplot [blue,dashed,line legend, sharp plot,update limits=false] coordinates { (-200,77798.0) (2000,77798.0) };
		\end{axis}
		\begin{axis}[name=plot2,at={($(plot1.east)+(1.1cm,0cm)$)},anchor=west,height=4cm,width=3.8cm,ylabel={depth},xlabel={\#ancillary qubits$(\ell)$},legend cell align=left,legend pos=outer north east]

			\addplot [blue,line legend, sharp plot,update limits=false,mark=diamond*]coordinates{( 100 , 518073 )};
			\addplot [blue,line legend, sharp plot,update limits=false,mark=triangle*]coordinates{( 100 , 518073 )};
			\addplot [blue,dashed,line legend, sharp plot,update limits=false,mark=square*,every mark/.append style={solid, fill=white}]coordinates{( 100 , 518073 )};
			\addplot [blue,dashed,line legend, sharp plot,update limits=false,mark=star]coordinates{( 100 , 518073 )};
			\addplot [mark=square*] coordinates{
			    ( 100 , 518073 )
			    ( 150 , 192137 )
			    ( 200 , 115733.8 )
			    ( 250 , 97113.8 )
			    ( 300 , 87633.8 )
			    ( 350 , 81039.4 )
				( 400 , 75945.8 )
				( 800 , 51419.4 )
				( 1200 , 42701.8 )
				( 1600 , 36413.8 )
				( 2000 , 32925 )
				( 2400 , 31275.4 )
				( 2800 , 29334.6 )
				( 3200 , 27402.6 )
				( 3600 , 27085.8 )
				( 4000 , 25232.2 )
				( 4400 , 25236.2 )
				( 4800 , 25048.2 )
				( 5200 , 24280.2 )
				( 5600 , 23053.8 )
				( 6000 , 23028.2 )
				( 6400 , 22854.6 )
				( 6800 , 22714.6 )
				( 7200 , 22330.6 )
				( 7600 , 22041.8 )
				( 7943 , 20871.4 )
			};
			\addplot [mark=*] coordinates{

			    ( 200 , 416178.6 )
			    ( 300 , 157496.2 )
			    ( 400 , 122659.4 )
			    ( 500 , 104660.2 )
			    ( 600 , 97362.6 )
			    ( 700 , 90723.4 )
				( 800 , 83001 )
				( 1600 , 55252.2 )
				( 2400 , 45832.2 )
				( 3200 , 38921 )
				( 4000 , 35056.2 )
				( 4800 , 33095.4 )
				( 5600 , 31045 )
				( 6400 , 28857.8 )
				( 7200 , 28613.8 )
				( 8000 , 26449 )
				( 8800 , 26475.4 )
				( 9600 , 26225.8 )
				( 10400 , 25390.6 )
				( 11200 , 24032.2 )
				( 12000 , 23855.4 )
				( 12800 , 23816.2 )
				( 13600 , 23793.8 )
				( 14400 , 23393 )
				( 15200 , 23025.8 )
				( 15887 , 21734.6 )
			};
			\addplot[black,densely dotted, every mark/.append style={solid, fill=white},mark=*] coordinates{
				( 7943 , 389242 )
			};
			\addplot[black,densely dotted, every mark/.append style={solid, fill=white},mark=otimes*] coordinates{
				( 15887 , 778498 )
			};
			\addplot [black,densely dotted,line legend, sharp plot,update limits=false] coordinates { (-2000,389242) (20000,389242) };
			\addplot [black,densely dotted,line legend, sharp plot,update limits=false] coordinates { (-2000,778498) (20000,778498) };
			\legend{Alg.2: $n=40$,Alg.2: $n=80$,Qiskit: $n=40$,Qiskit: $n=80$,Alg.2: $n=400$,Alg.2: $n=800$,Qiskit: $n=400$,Qiskit: $n=800$}
		\end{axis}
	\end{tikzpicture}
		\centering
		\caption{The change of the depth of circuits synthesized by the depth-oriented algorithm.}
	\end{subfigure}
\caption{The performance of algorithms. Choose the random 4-CNF with 40, 80, 400, and 800 variables to compare the performance of our algorithm and qiskit's. The result of qiskit contains only one point.
}
\label{fig:fig_our_result}
\end{figure}
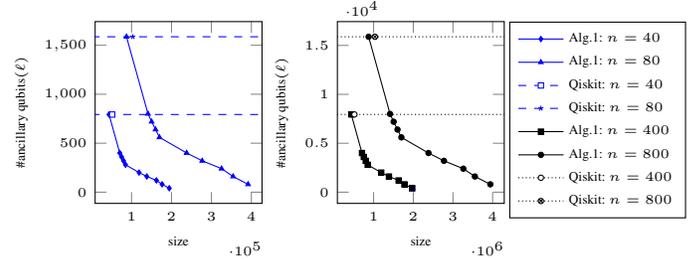
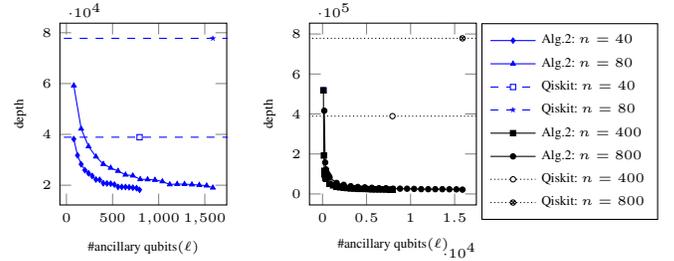

{In Figure~\ref{fig:fig_our_result} (a), we compare the size of the quantum circuit synthesized by our algorithm and the $\CNF$ synthesis algorithm used in the qiskit.} 
The circuit synthesis algorithm in qiskit requires $2m-1$ ancillary qubits, so there is only one point and the corresponding horizontal dashed line in Figure ~\ref{fig:fig_our_result}~\cite{qiskit_alg}. This Figure shows that when $2m-1$ ancillary qubits are used, the size-oriented algorithm can generate a quantum circuit with a smaller size than qiskit. If we want to reduce the number of ancillary qubits significantly, the corresponding size will only increase by a constant multiple. For example, for a random 4-$\CNF$ with $n$ equal to 80, qiskit needs 1587 ancillary qubits to synthesize a quantum circuit of size 103205, while the size-oriented algorithm can use 80 ancillary qubits {(about $5\%$ of qiskit)} to synthesize a circuit of size 391760 {(less than 4 times)}.


In Figure ~\ref{fig:fig_our_result}~(b), a more notable advantage appears in comparing the circuit depth between the depth-oriented algorithm and the algorithm used in qiskit. The depth-oriented algorithm requires very few ancillary qubits to synthesize low-depth quantum circuits. 
When using the same $2m-1$ ancillary qubits as qiskit, the circuit depth synthesized by depth-oriented algorithm is significantly lower. 
For example, for a random 4-$\CNF$ with $n$ equal to 800, qiskit needs 15887 ancillary qubits to synthesize a quantum circuit of depth 778498. In comparison, the depth-oriented algorithm can use 200 ancillary qubits { (about $1.2\%$ of qiskit)} to synthesize a circuit of depth 416179 { (about $53.5\%$ of qiskit)}. If the depth-oriented algorithm uses 15887 ancillary qubits, the depth of the circuit can be reduced to 21735 { (about $2.8\%$ of qiskit)}. {For both two algorithms, the quantum cost of the output circuit is declined with the growth of the number of ancillary qubits.} Despite a few single points, our experimental results are in good agreement with the theory.

\begin{table}[htb]
    \footnotesize
	\centering
	\begin{tabular}{|cccc||ll|ll|}
		\toprule[0.5pt]
		\hline
		&&&&\multicolumn{2}{c|}{size (Alg.1)} & \multicolumn{2}{c|}{depth (Alg.2)} \\
		$k$&$n$ & \#clause & \#ancillae & full round & one round& full round & one round \\ \hline
		3 & 40 & 170 & 240 & $1.8\times 10^{10}$ & 21384 & $6.2 \times 10^9$ & 7523 \\
        3 & 80 & 341 & 240 & $4.3\times 10^{16}$ & 49522 & $1.0 \times 10^{16}$ & 11586 \\
        5 & 40 & 844 & 240 & $4.3\times 10^{11}$ & 5.2 $\times 10^5$ & $7.6 \times 10^{10}$ & $92444$ \\
        5 & 80 & 1689 & 240 & $1.0\times 10^{18}$ & $1.2 \times 10^6$ & $1.1 \times 10^{17}$ & $1.3 \times 10^5$ \\
        7 & 40 & 3511 & 240 & $3.4\times 10^{12}$ & $4.1 \times 10^6$ & $6.7 \times 10^{11}$ & $8.1 \times 10^5$ \\
        7 & 80 & 7023 & 240 & $7.3\times 10^{18}$ & $8.4 \times 10^6$ & $9.3 \times 10^{17}$ & $1.1 \times 10^6$ \\
		\hline
		\bottomrule[0.5pt]
	\end{tabular}
	\caption{The quantum resource estimation for solving $k$-SAT problems using Grover's algorithm, where $k=3,5,7$. $n$ is the number of variables. The left 4 columns are the parameters of the $k$-SAT problem instance, and the rest columns show the quantum cost needed to solve the $k$-SAT problem via Grover's algorithm and our synthesis algorithms. We list the full round cost and the one round cost separately.}
	\label{tab:sat}  
\end{table}

Further, we apply our synthesis algorithms to estimate the quantum resources required for solving $k$-SAT using Grover's algorithm.
In \cite{campbell2019applying}, they also estimate the quantum resources required for solving $14$-SAT algorithm. To solve a $65$ to $78$ variables $14$-SAT,
the number of ancillary qubits used is $10^{12}$  to $10^{14}$, which is unavailable in NISQ era. Hence, we fix the number of ancillary qubits as $240$, which is more available in NISQ era.
In Grover's algorithm there is multiple rounds of Grover iteration. Table~\ref{tab:sat} shows the quantum resources needed for both the full round (reaching the highest success probability) and one round execution of Grover iteration. 
For example, a random $7$-SAT with 80 variables, which is among the most challenging instances that a classical computer can solve today~\cite{sc2013,sc2014,sc2016,sc2017,sc2018}, can be solved by using 240 ancillary qubits and a $7.3 \times 10^{18}$-size quantum circuit via Grover's algorithm. 

\section{Discussion}
In this manuscript we design two synthesis algorithms for the different quantum costs. 
We first construct a general $p$-AND circuit. We design the size-oriented algorithm, which recursively use $p$-AND module to construct the circuit for SAT-oracle. Notice that the ancillary qubits used here could be dirty, which means this algorithm can use the temporarily idle qubits. The size of the circuit generated with this algorithm is $O(n(km/n)^{1+c})$, where $c=o(1)$ is determined by $\ell$.
Specially we can reduce the number of ancillary qubits to $O(\sqrt{m})$ with a constant ratio increase in the size of the circuit. 
We also prove a matched lower bound of this problem using counting method. 
Further, we propose the depth-oriented algorithm to reduce the depth of the quantum circuit. 
The depth of the circuit generate by depth-oriented algorithm is $O(k\log \ell (mS/\ell)^{1+c'})$, where $S=\max\{k/\log \ell,1\}$ and $c'=o(1)$ is determined by $\ell$. We design several experiments to evaluate the performance of our algorithm. Finally, we apply our synthesis algorithms to give an estimate of quantum costs to solve the SAT problem. 

Some interesting open problems left. Can we use the dirty ancillary qubits to replace the clean ancillary qubits in the synthesis algorithm? Is there some essential difference between the clean and dirty ancillary qubits in general circuit? Are there some efficient algorithms that can generate size-optimal or depth-optimal circuit for any given $\CNF_{n,m}^k$ instance?
\bibliographystyle{unsrt}
\bibliography{ref}

\begin{thebibliography}{10}
\bibitem{CNFintro}
\rm{CNF} is an \rm{AND} of several clauses, each clause is an \rm{OR} of the
  variables or their negations. \text{T}he number of variables is usually
  denoted by $n$ and the number of clauses is denoted by $m$. \text{T}he number
  of variables used in a single clause is called the width of the clauses, and
  the width of \text{CNF} is defined by the max width of the clauses, which is
  denoted by $k$.
\bibitem{qiskit_alg}
\text{T}he synthesis algorithm used in qiskit has been changed. \text{W}hen we
  wrote this manuscript, the algorithm used in qiskit is first generating all
  clauses in the clean ancillary qubits and then merged them up. \text{T}he new
  algorithm used in qiskit is a heuristic algorithm, which can not work when
  the input $n$ is large. \text{H}ence, we still compare the result with the
  old synthesis algorithm in qiskit.
\end{thebibliography}
\end{document}


\title{Supplementary materials for ``
Efficient quantum circuit synthesis for SAT-oracle with limited ancillary qubit
''
}
\author{Shuai Yang}
\affiliation{Institute of Computing Technology, Chinese Academy of Sciences, 100190 Beijing, China}
\affiliation{University of Chinese Academy of Sciences, 100049 Beijing, China}

\author{Wei Zi}
\affiliation{Institute of Computing Technology, Chinese Academy of Sciences, 100190 Beijing, China}
\affiliation{University of Chinese Academy of Sciences, 100049 Beijing, China}

\author{Bujiao Wu}
\affiliation{Center on Frontiers of Computing Studies, Peking University, Beijing 100871, China}

\author{Cheng Guo}
\affiliation{Institute of Computing Technology, Chinese Academy of Sciences, 100190 Beijing, China}
\affiliation{University of Chinese Academy of Sciences, 100049 Beijing, China}

\author{Jialin Zhang}
\affiliation{Institute of Computing Technology, Chinese Academy of Sciences, 100190 Beijing, China}
\affiliation{University of Chinese Academy of Sciences, 100049 Beijing, China}

\author{Xiaoming Sun\footnote{sunxiaoming@ict.ac.cn}}
\affiliation{Institute of Computing Technology, Chinese Academy of Sciences, 100190 Beijing, China}
\affiliation{University of Chinese Academy of Sciences, 100049 Beijing, China}
\email[]{sunxiaoming@ict.ac.cn}

\maketitle
\section{Notation and preliminary}
A Boolean formula consists of the variables and logical operations AND, OR, and NOT. In specific, 
 a Conjunction Normal Form (Disjunctive Normal Form) formula is the AND (OR) of OR's (AND's) of variables or their negations. To synthesis a CNF (DNF) formula $f$ is to find a quantum circuit $C_f$ such that for any input $x$, we have $C_f\ket{x}\ket{y}\to \ket{x}\ket{y\oplus f(x)}.$ X gate, Controlled-NOT gate (CNOT gate), and Toffoli gate form a universal gate set for Boolean functions oracle.
 A layer in a quantum circuit is a set of consecutive disjoint quantum gates. We use the number of the elementary gates (CNOT gate and single-qubit gate) and the number of layers to measure the cost of a quantum circuit.

For convenience, we use the notation $\CNF_{n,m}^k$ to denote an instance of $n$ variables $m$ clauses $k$-CNF and $\text{Size}_\ell (\CNF_{n,m}^k)$ to denote the size of the circuit generated by the first algorithm for $\CNF_{n,m}^k$ with $\ell$ ancillary qubits. We may omit the rounding notation `$\lfloor \rfloor$' for convenience in the manuscript.

A multi-controlled Toffoli gate in quantum circuit can calculate AND/OR of variables. Denote $Tof_{q_1,q_2,\dots,q_k}^{q_t}$ as the $k$-controlled Toffoli gate, where the control qubits are $\{q_1,q_2,\cdots,q_k\}$ and the target qubit is $q_t$. For examples:
\begin{align*}
    Tof_{q_{1},q_{2},\cdots,q_{k}}^{q_t}\ket{q}\ket{c}\to\ket{q}\ket{c\oplus\wedge_{j=1}^k q_{j}}, \\
    (\otimes_{j=1}^k X_{j}) X_{q_t}(Tof_{q_{1},q_{2},\cdots,q_{k}}^{q_t}) (\otimes_{j=1}^k X_{i_j})\ket{q}\ket{c}\\
    \to\ket{q}\ket{c\oplus\vee_{j=1}^k q_{j}}.
\end{align*}

\section{The detail of General AND (OR) circuit}
\label{sec:dirty_anci}
\begin{lemma}[general $p$-AND circuit]
For any natural number $p$, general $p$-AND circuit can be implemented with $2p-2$ {dirty} ancillary qubits, $O(p)$ Toffoli gate and 4 calls of {each} $\Ocal_i$.
\label{lem:dirty}
\end{lemma}

For convenience, let us introduce some notations.
\begin{equation*}
    q'_{i} = \left\{
    \begin{aligned}
        q_i \oplus \left(q_{i-1}\wedge g_{i-p+1}(x)\right)& & 2p-2\ge i \ge p+2\\
        q_i\oplus\left(q_{1}\wedge g_{2}(x)\right)& & i=p+1\\
        q_i \oplus g_{i}(x)& & p\ge i\ge 1\\
    \end{aligned}
    \right.
\end{equation*}
and $q''_{i} = q_{i}\oplus (\wedge_{k=1}^{i-p+1}g_{k}(x)),~i \in \{p+1,p+2\cdots 2p-2\}$.
Let $q_t$ be the target qubit,  $q_t^{a}=q_t\oplus(\bigwedge_{k=1}^{a}g_{k}(x))$, and $q'_t=q_{t}\oplus (q_{p}\wedge g_p(x)).$
Let $Q_{a,b}=\otimes_{k=a}^{b}\ket{q_k}$, $Q'_{a,b}=\otimes_{k=a}^{b}\ket{q'_k}$, and $Q''_{a,b}=\otimes_{k=a}^{b}\ket{q''_k}$.
The number of dirty
ancillary qubits $\ell=2p-2.$
\begin{proof}
Using Toffoli gate can easily merge the information stored in the qubits. However if the information is stored in the oracle, we need to apply the oracle in some qubits at first. The information stored in the qubits will influence the result. Hence, we add additional operation to eliminate the dirty information. 
We firstly divide the $\mathcal{C}$ into 2 sub-circuit:  merge stage $\mathcal{C}_2$ and restore stage $\mathcal{C}_1$.
The construct of quantum circuit $\mathcal{C}_1$ contains $2p-3$ steps.  These steps can be divided into 3 phases : \emph{Up phase}, \emph{Top phase} and \emph{Down phase}. The \emph{Up phase} first store the information about $q_{p+1-i}$ and oracle $\Ocal_{p+1-i}$ at qubit $q_{2p-1-i}$.
The \emph{Top phase} then merge the information about $\Ocal_{1}$ and $\Ocal{2}$.
In the \emph{Down phase}, we merge the all the information store in the $\Ocal_i$. Notice the dirty information is added twice under $\Fbb_2$.
Finally in $\Ccal_2$ we repeat step 2 to step $2p-4$ to restore the ancillary qubits.

\begin{enumerate}
    \item \emph{Up phase}: In this phase, we add the information in the ancillary qubits to corresponding qubits, which can help to eliminate unexpected information. There are 3 parts in step $i\in [p-2]$. The details of each part are shown as follows.
    \begin{enumerate}
        \item In step $i.1$, as well as step $i.3$, corresponds to a Toffoli gate. The control qubits are $q_{p+1-i}$ and $q_{2p-1-i}$. When $i=1$, the target qubit is $q_t$, otherwise the target is $q_{2p-i}$.
        \item In step $i.2$, we call the $\Ocal_{p+1-i}$ at $q_{p+1-i}$.
    \end{enumerate}
    
    At the begining, the state is at the 
    \[
        Q_{1,\ell}\ket{q_t}
    \]
    After step 1.1, we apply an Toffoli gate to store the dirty information stored in the dirty ancillary qubits $q_{p+1}$ and $q_{2p}$. Then the state transfer to 
    \[
        Q_{1,\ell}\ket{q_t\oplus(q_{p}\wedge q_{2p-2})}.
    \]
    Then, we apply an $\Ocal_{p+1}$ at the qubit $q_{p+1}$. This step add the information of the oracle into the qubits. The state after step 1.2 becomes
    \[
        Q_{1,p-1}Q'_{p,p}Q_{p+1,2p-2}\ket{q_t\oplus(q_{p}\wedge q_{2p-2})}.
    \]
    Finally in step 1.3, we use a Toffoli gate to add the information to the target qubit and some dirty information has been eliminated. The state after step 1 is
    \[
        Q_{1,p-1}Q'_{p,p}Q_{p+1,2p-2}\ket{q'_t}.
    \]
    
    Same to the analyze in the step 1, the state before the step $i.1$ is
    \[
    Q_{1,p-i+1}Q'_{p-i+2,p}Q_{p+1,2p-i}Q'_{2p-i+1,2p-2}\ket{q'_{t}}.
    \]
    After step $i.3$, the state transfer to 
    \[
        Q_{1,p-i}Q'_{p-i+1,p}Q_{p+1,2p-i-1}Q'_{2p-i,2p-2}\ket{q'_{t}}.
    \]
    \item \emph{Top phase}: In this phase, we implement a circuit that merges two clauses and stores the result in an ancillary qubit. There are seven parts in step $p-1$. The details of each part are shown as follows.
    \begin{enumerate}
        \item In steps $(p-1).j$, $j\in\{1,3,5,7\}$, the operation are the same. Each step corresponds to a Toffoli gate. The control qubits of Toffoli gate is $q_1$ and $q_2$, and the target qubit is $q_{p+1}.$
        \item In steps $(p-1).2$, as well as $p.6$, we call the $\Ocal_{2}$ at $q_{2}$.
        \item In steps $(p-1).4$, we call the $\Ocal_{1}$ at $q_{1}$.
    \end{enumerate}
    
    The step from $(p-1).1$ to $(p-1).3$ are similar to the step in \emph{Up phase}. The state after step $(p-1).3$ is 
    \[
        \ket{q_1}Q'_{2,2p-2}\ket{q'_t}
    \]
    The step $(p-1).4$ just apply an Oracle $\Ocal_{1}$. So the state is $Q'_{1,2p-2}\ket{q'_t}$.
    
    The step from $(p-1).5$ to $(p-1).7$ restore the qubit $2$ and store the $g_1\wedge g_2(x)$ without dirty information in qubit $p+1$. The equation below only focus on the qubits $q_1,q_2$ and $q_{p+2}$.
    \begin{align*}
        &\ket{q'_1}\ket{q'_2}\ket{q'_{p+1}}\\
        \xrightarrow{(p-1).5}&\ket{q'_1}\ket{q'_2}\ket{q'_{p+1}\oplus ((q_1\oplus g_1(x))\wedge (q_2\oplus g_2(x)))}\\
        \xrightarrow{(p-1).6}&\ket{q'_1}\ket{q_2}\ket{q'_{p+1}\oplus ((q_1\oplus g_1(x))\wedge (q_2\oplus g_2(x)))}\\
        \xrightarrow{(p-1).7}&\ket{q'_1}\ket{q_2}\ket{q'_{p+1}\oplus ((q_1\oplus g_1(x))\wedge g_2(x))}\\
        =&\ket{q'_1}\ket{q_2}\ket{q_{p+1}\oplus(g_1(x)\wedge g_2(x)).}\\
    \end{align*}
    \item \emph{Down phase}: In this phase, we merge all the clauses in the ancillary qubits. With the help of \emph{Up phase}, the target qubit stores the value of input CNF. There are 3 parts in step $i\in\{p,p+2,\cdots,2p-3\}$ .
    
    \begin{enumerate}
        \item Step i.1, as well as step i.3, corresponds to a Toffoli gate. The control qubits are $q_{i - p+3}$ and $q_{i+1}$. When $i=2p-3$, the target qubit is $q_t$, otherwise the target is $q_{i+2}$.
        \item In step i.2, we call the $\Ocal_{i-p+3}$ at $q_{i-p+3}$.
    \end{enumerate}
    The analyse is similar to the step $(p-1).5$ to $(p-1).7$. In each step, we restore a qubit $i-p+3$ and store the correct information in qubit $q_{i+3}$. 
\end{enumerate}

After the $\Ccal_1$, the state is 
\[
    \ket{q'_1(x)}Q_{2,p}Q''_{p+1,2p-2}\ket{q_t\oplus f(x)}.
\]
What we need to do is repeating the steps from $2$ to $2p-4$ to restore the ancillary qubits.

Totally, we use at most $8p-12$ Toffoli gates and call each $\Ocal_i$ at most four times. In the merge stage, each step except step $p$ contain 2 Toffoli gates. The total number of Toffoli gates in merge stage is $2(2p-3-1)+4=4p-4$. Similarly, we use $4p-8$ Toffoli gates in restore stage. So we use at most $8p-12$ Toffoli gates in $p$-AND circuit. In each stage, each oracle $\Ocal_i$ is called at most 2 times. So each $\Ocal_i$ is called at most 4 times in the $p$-AND circuit. 
\end{proof}

\begin{corollary}
By adding X gates when we call the $\Ocal$ and adding X gates at target qubit, we can construct a circuit $\mathcal{C}$ for function $f(x)=\bigvee g_i(x)$.
\end{corollary}

\section{Size-oriented synthesis algorithm}


This section will introduce how to use the general $p$-AND circuit to construct the circuit for SAT-Oracle.

\begin{figure*}[ht]
\Qcircuit @C=0.5em @R=.5em{
\lstick{}&&&&&&&&\textbf{Merge stage}&&&&&&&&&&&&&\textbf{Restore stage}\\
\\
\lstick{\ket{x}} & \qw &\qw{/} &\multigate{5}{\mathcal{O}_3^{q_3}}&\qw
&\qw&\multigate{5}{\mathcal{O}_2^{q_2}}&\qw&\multigate{5}{\mathcal{O}_1^{q_1}}&\qw&\multigate{5}{\mathcal{O}_2^{q_2}}&\qw
&\qw&\multigate{5}{\mathcal{O}_3^{q_3}}&\qw&\qw&\qw \gategroup{3}{3}{8}{15}{1.2em}{--}  %
&\qw&\qw&\multigate{5}{\mathcal{O}_2^{q_2}}&\qw&\qw&\multigate{5}{\mathcal{O}_1^{q_1}}&\qw&\multigate{5}{\mathcal{O}_2^{q_2}}&\qw&\qw&
&&\multigate{5}{\mbox{3-AND}}&\qw&\ket{x}  \gategroup{3}{18}{8}{26}{1.2em}{--} \\
\lstick{\ket{q_1}} &\qw&\qw &\ghost{\mathcal{O}_3^{q_3}}&\qw
&\ctrl{1}&\ghost{\mathcal{O}_2^{q_2}}&\ctrl{1}&\ghost{\mathcal{O}_1^{q_1}}&\ctrl{1}&\ghost{\mathcal{O}_2^{q_2}}&\ctrl{1}
&\qw &\ghost{\mathcal{O}_3^{q_3}} &\qw&\qw&\qw%
&\qw&\ctrl{1}&\ghost{\mathcal{O}_2^{q_2}}&\ctrl{1}&\qw&\ghost{\mathcal{O}_1^{q_1}}&\ctrl{1}&\ghost{\mathcal{O}_2^{q_2}}&\ctrl{1}&\qw&
&&\ghost{\mbox{3-AND}}&\qw&~\ket{q_1}\\
\lstick{\ket{q_2}} &\qw&\qw&\ghost{\mathcal{O}_3^{q_3}}&\qw
&\ctrl{2}&\ghost{\mathcal{O}_2^{q_2}}&\ctrl{2}&\ghost{\mathcal{O}_1^{q_1}}&\ctrl{2}&\ghost{\mathcal{O}_2^{q_2}}&\ctrl{2}
&\qw&\ghost{\mathcal{O}_3^{q_3}} &\qw&\qw&\qw%
&\qw&\ctrl{2}&\ghost{\mathcal{O}_2^{q_2}}&\ctrl{2}&\qw&\ghost{\mathcal{O}_1^{q_1}}&\ctrl{2}&\ghost{\mathcal{O}_2^{q_2}}&\ctrl{2}&\qw&=
&&\ghost{\mbox{3-AND}}&\qw&~\ket{q_2}\\
\lstick{\ket{q_3}} &\qw&\ctrl{1}&\ghost{\mathcal{O}_3^{q_3}}&\ctrl{1}
&\qw&\ghost{\mathcal{O}_2^{q_2}}&\qw&\ghost{\mathcal{O}_1^{q_1}}&\qw&\ghost{\mathcal{O}_2^{q_2}}&\qw
&\ctrl{1}&\ghost{\mathcal{O}_3^{q_3}}&\ctrl{1}&\qw&\qw%
&\qw&\qw&\ghost{\mathcal{O}_2^{q_2}}&\qw&\qw&\ghost{\mathcal{O}_1^{q_1}}&\qw&\ghost{\mathcal{O}_2^{q_2}}&\qw&\qw&
&&\ghost{\mbox{3-AND}}&\qw&~\ket{q_3}\\
\lstick{\ket{q_4}} &\qw &\ctrl{1}&\ghost{\mathcal{O}_3^{q_3}}&\ctrl{1}
&\targ&\ghost{\mathcal{O}_2^{q_2}}&\targ&\ghost{\mathcal{O}_1^{q_1}}&\targ&\ghost{\mathcal{O}_2^{q_2}}&\targ
&\ctrl{1}&\ghost{\mathcal{O}_3^{q_3}}&\ctrl{1}&\qw&\qw%
&\qw&\targ&\ghost{\mathcal{O}_2^{q_2}}&\targ&\qw&\ghost{\mathcal{O}_1^{q_1}}&\targ&\ghost{\mathcal{O}_2^{q_2}}&\targ&\qw&
&&\ghost{\mbox{3-AND}}&\qw&~\ket{q_4}\\
\lstick{\ket{q_t}} &\qw &\targ&\ghost{\mathcal{O}_3^{q_3}}&\targ
&\qw&\ghost{\mathcal{O}_2^{q_2}}&\qw&\ghost{\mathcal{O}_1^{q_1}}&\qw&\ghost{\mathcal{O}_2^{q_2}}&\qw
&\targ&\ghost{\mathcal{O}_3^{q_3}}&\targ&\qw&\qw%
&\qw&\qw&\ghost{\mathcal{O}_2^{q_2}}&\qw&\qw&\ghost{\mathcal{O}_1^{q_1}}&\qw&\ghost{\mathcal{O}_2^{q_2}}&\qw&\qw&
&&\ghost{\mbox{3-AND}}&\qw&~~~~~~~~~~~~~~~~~~~~\ket{q_t\oplus \wedge_1^3 g_i(x)}\\
\\
\\
\textbf{Step:}~~~~&&\textbf{\scriptsize1.1}&\textbf{\scriptsize1.2}&\textbf{\scriptsize1.3}&~\textbf{\scriptsize2.1}&\textbf{\scriptsize2.2}&\textbf{\scriptsize2.3}&\textbf{\scriptsize2.4}&\textbf{\scriptsize2.5}&\textbf{\scriptsize2.6}&\textbf{\scriptsize2.7}&~\textbf{\scriptsize3.1}&\textbf{\scriptsize3.2}&\textbf{\scriptsize3.3}&&&&~\textbf{\scriptsize2.1}&\textbf{\scriptsize2.2}&\textbf{\scriptsize2.3}&&\textbf{\scriptsize2.4}&\textbf{\scriptsize2.5}&\textbf{\scriptsize2.6}&\textbf{\scriptsize2.7}\\
\\
\textbf{Phase:}~~~~~&&&\textbf{Up phase}&&&&&\textbf{Top phase}&&&&&\textbf{Down phase}&&&&&&&&&\textbf{Top phase}\\
}
\caption{A example of general $3$-AND circuit.}
\label{fig:mer_cir}
\end{figure*}
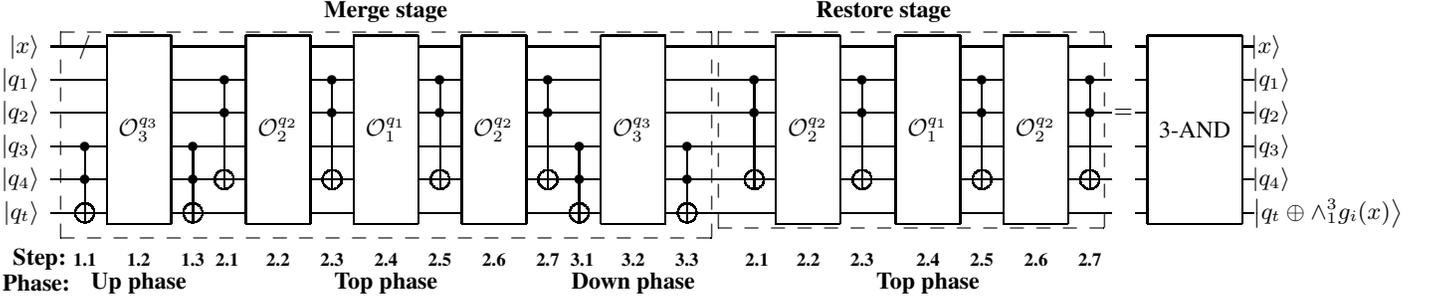

\begin{lemma}
Any $n$ variables $m$ clauses $k$-CNF $f\in \CNF_{n,m}^k$ can be implemented by $O(km^{1+\log_{\ell/2+1} 4})$-size quantum circuits  $\mathcal{C}$ with $\ell$ ancillary qubits. 
\end{lemma}
\begin{proof}
We divide the clauses in CNF into $p=\lfloor\ell/2\rfloor+1$ sub-blocks recursively until in each blocks only one clause or less. An oracle $\Ocal$ for a single clause can be realized by a $k$-controlled Toffoli gate and several X gate, where $k$ is the width of the clause, which means $\text{Size}_0(\CNF_{n,1}^{k})=O(k)$.

Lemma \ref{lem:dirty} shows that we can construct a circuit to calculate the AND(OR) of some sub-function. By recursively using the general $p$-AND/OR circuit, we can finally construct the circuit for any given CNF. In each recursion we need $O(\ell)$ elementary gates and call $4p$ sub-blocks oracle.

The total quantum cost is $\text{Size}_{\ell}(\CNF_{n,m}^{k}) = 4p \text{Size}_{\ell}(\CNF_{n,m/p }^{k}) + 4\ell.$ Notice that $\text{Size}_\ell(\CNF_{n,1}^{k})\le\text{Size}_0(\CNF_{n,1}^{k})=O(k).$
Solving this recursion formula, we have that $\text{Size}_{\ell}(\text{CNF}_{n,m}^{k})=O(km^{1+\log_{\ell/2+1} 4}).$
\end{proof}

Ancillary qubits used in this section are dirty ancillary qubits, which means we can use the unused input qubits as ancillary qubits. In a $t~ (t\!\le\!\frac{n-n^\epsilon}{k})$ clauses CNF formula, at most $kt$ variables is used in this CNF formula.
To generate such a CNF formula, we can regard other input qubits as ancillary qubits, which means 
$\text{Size}_0\left(\CNF_{n,\frac{n-n^\epsilon}{k}}^k\right)=\text{Size}_{n^\epsilon}\left(\CNF_{n,\frac{n-n^\epsilon}{k}}^k\right)=O(n)$. This idea can improve the upper bound that $\text{Size}_{\ell}(\text{CNF}_{n,m}^{k})=O\left(n\left(\frac{km}{n-n^\epsilon}\right)^{1+\log_{\ell+1} 4}\right)$. Then the upper bound is proved.

If the ancillary qubits are clean, which means the initial state of ancillary qubits are $\ket{0}$ at the beginning, the outer-most recursive circuit can be improved. 
We can merge $\ell$ terms into one in the outer-most recursion. This optimization will reduce the quantum cost by about half.
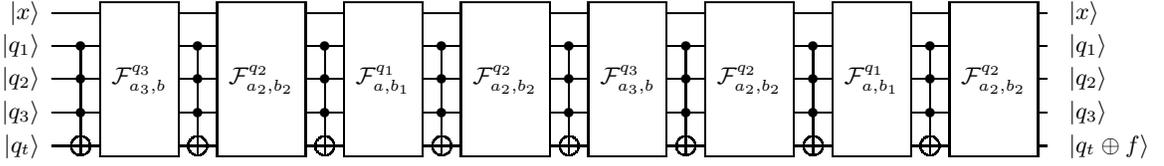
\begin{figure*}[htb]
\centerline{
\Qcircuit @C=0.4em @R=.5em{
\lstick{\ket{x}}&\qw&\qw&\multigate{4}{\mathcal{F}_{a_3,b}^{q_3}}&\qw&\multigate{4}{\mathcal{F}_{a_2,b_2}^{q_2}}&\qw&\multigate{4}{\mathcal{F}_{a,b_1}^{q_1}}&\qw&\multigate{4}{\mathcal{F}_{a_2,b_2}^{q_2}}&\qw&\multigate{4}{\mathcal{F}_{a_3,b}^{q_3}}&\qw&\multigate{4}{\mathcal{F}_{a_2,b_2}^{q_2}}&\qw&\multigate{4}{\mathcal{F}_{a,b_1}^{q_1}}&\qw&\multigate{4}{\mathcal{F}_{a_2,b_2}^{q_2}}&\qw&\rstick{\ket{x}}\\
\lstick{\ket{q_1}}&\qw&\ctrl{1}&\ghost{\mathcal{F}_{a_3,b}^{q_3}}&\ctrl{1}&\ghost{\mathcal{F}_{a_2,b_2}^{q_2}}&\ctrl{1}&\ghost{\mathcal{F}_{a,b_1}^{q_1}}&\ctrl{1}&\ghost{\mathcal{F}_{a_2,b_2}^{q_2}}&\ctrl{1}&\ghost{\mathcal{F}_{a_3,b}^{q_3}}&\ctrl{1}&\ghost{\mathcal{F}_{a_2,b_2}^{q_2}}&\ctrl{1}&\ghost{\mathcal{F}_{a,b_1}^{q_1}}&\ctrl{1}&\ghost{\mathcal{F}_{a_2,b_2}^{q_2}}&\qw&\rstick{\ket{q_1}}\\
\lstick{\ket{q_2}}&\qw&\ctrl{1}&\ghost{\mathcal{F}_{a_3,b}^{q_3}}&\ctrl{1}&\ghost{\mathcal{F}_{a_2,b_2}^{q_2}}&\ctrl{1}&\ghost{\mathcal{F}_{a,b_1}^{q_1}}&\ctrl{1}&\ghost{\mathcal{F}_{a_2,b_2}^{q_2}}&\ctrl{1}&\ghost{\mathcal{F}_{a_3,b}^{q_3}}&\ctrl{1}&\ghost{\mathcal{F}_{a_2,b_2}^{q_2}}&\ctrl{1}&\ghost{\mathcal{F}_{a,b_1}^{q_1}}&\ctrl{1}&\ghost{\mathcal{F}_{a_2,b_2}^{q_2}}&\qw&\rstick{\ket{q_2}}\\
\lstick{\ket{q_3}}&\qw&\ctrl{1}&\ghost{\mathcal{F}_{a_3,b}^{q_3}}&\ctrl{1}&\ghost{\mathcal{F}_{a_2,b_2}^{q_2}}&\ctrl{1}&\ghost{\mathcal{F}_{a,b_1}^{q_1}}&\ctrl{1}&\ghost{\mathcal{F}_{a_2,b_2}^{q_2}}&\ctrl{1}&\ghost{\mathcal{F}_{a_3,b}^{q_3}}&\ctrl{1}&\ghost{\mathcal{F}_{a_2,b_2}^{q_2}}&\ctrl{1}&\ghost{\mathcal{F}_{a,b_1}^{q_1}}&\ctrl{1}&\ghost{\mathcal{F}_{a_2,b_2}^{q_2}}&\qw&\rstick{\ket{q_3}}\\
\lstick{\ket{q_t}}&\qw&\targ&\ghost{\mathcal{F}_{a_3,b}^{q_3}}&\targ&\ghost{\mathcal{F}_{a_2,b_2}^{q_2}}&\targ&\ghost{\mathcal{F}_{a,b_1}^{q_1}}&\targ&\ghost{\mathcal{F}_{a_2,b_2}^{q_2}}&\targ&\ghost{\mathcal{F}_{a_3,b}^{q_3}}&\targ&\ghost{\mathcal{F}_{a_2,b_2}^{q_2}}&\targ&\ghost{\mathcal{F}_{a,b_1}^{q_1}}&\targ&\ghost{\mathcal{F}_{a_2,b_2}^{q_2}}&\qw&\rstick{\ket{q_t\oplus f}}\\
}
}
\caption{When the number of ancillary qubits is small, another synthesis circuit for the CNF formula. Here $f\in \CNF_{n,(b-a)}^k.$}
\label{fig:small_anci}
\end{figure*}

Approximate Toffoli gate consists of 3 CNOT gates and four single-qubit gates. Like the Toffoli gate, approximate Toffoli gate can implement AND/OR gate in the quantum circuit, with an additional change over control qubit. Using approximate Toffoli to synthesize Toffoli gate can reduce the quantum cost of $m$-control Toffoli gate. 

When the number of ancillary qubits is small, we can use the circuit shown in Figure \ref{fig:small_anci} to replace the circuit in Figure \ref{fig:mer_cir}. The recursive circuit shown in Figure \ref{fig:small_anci} can merge $\ell$ functions, rather than $\ell/2$ functions shown in the previous discussion, in one recursion.

Choosing these optimizations can help us synthesize CNF with less quantum cost.

\section{Depth-oriented synthesis algorithm}


To further reduce the depth of the circuit, we need to parallelize our construct algorithm. In the size-oriented synthesis algorithm, each sub-block is implemented one by one. So we try to use some of the ancillary qubits, which are clean, to parallelize the circuit.

The framework of the depth-oriented synthesis algorithm is similar to the algorithm described in the size-oriented synthesis algorithm. Different from the size-oriented synthesis algorithm, we divide the ancillary qubits into 3 registers: $q_{mem}, q_{dirty}, q_{clean}.$ The size of these 3 registers are $\frac{(S-1)\ell}{S+1}, \frac{\ell}{S+1}, \frac{\ell}{S+1}$, where $S=\max\{\frac{k}{\log \ell},1\}.$ We run the recursive procedure by using $q_{dirty}$ as ancillary qubits. The only difference lies in the inner-most recursion of our algorithm. We use all the ancillary qubits to synthesize $\frac{\ell}{S+1}$ clauses in parallel. 

\begin{figure*}[ht]
    \centerline{
    \Qcircuit @C=0.5em @R=.5em{
    \lstick{\ket{x}}&\qw / &\multigate{1}{Copy}&\multigate{3}{Clause}&\qw&\multigate{4}{Reset}&\qw&\rstick{\ket{x}}\\
    \lstick{q_{mem},\ket{0}}&\qw / &\ghost{Copy}&\ghost{Clause}&\qw&\ghost{Reset}&\qw&\rstick{\ket{0}}\\
    \lstick{q_{dirty},\ket{q}}&\qw / &\qw&\ghost{Clause}&\qw&\ghost{Reset}&\qw&\rstick{\ket{q}}\\
    \lstick{q_{clean},\ket{0}}&\qw / &\qw&\ghost{Clause}&\multigate{1}{Merge}&\ghost{Reset}&\qw&\rstick{\ket{0}}\\
    \lstick{\ket{q_t}}&\qw&\qw&\qw&\ghost{Merge}&\ghost{Reset}&\qw&\rstick{\ket{q_t}}\\
    }}
	\caption{The inner-most recursion of algorithm. The ancillary qubits are divide into 3 registers: $q_{mem},q_{dirty},q_{clean}$, where the size of these registers are $\frac{(S-1)\ell}{S+1},\frac{\ell}{S+1},\frac{\ell}{S+1}$, respectively.}
	\label{fig:inner_recursion}
\end{figure*}
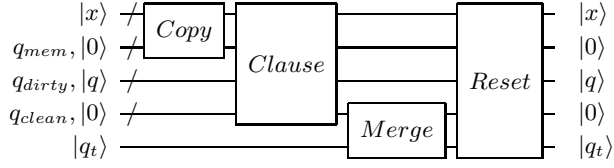

The inner-most recursion of our circuit is shown in Figure \ref{fig:inner_recursion}. There are four stages in the inner-most recursion: Copy stage, Clause stage, Merge stage, and Reset stage. 

For convenience, let $\mathcal{CO},\mathcal{CL},\mathcal{ME},\mathcal{RE}$ to denote the Copy stage, Clause stage, Merge stage and Reset stage, respectively. After these 4 stages, we synthesize a $f=\bigwedge_{j=1}^{{\lfloor\ell/S\rfloor}}C_j\in \CNF_{n,\lfloor\ell/S\rfloor}^k$  to the target qubit. Without loss of generality, let $\lfloor\ell/S\rfloor$ is a even.

In the Copy stage $\mathcal{CO}$, we copy the information of input qubits to the $q_{mem}$ register. 
\begin{align}
    \mathcal{CO}\ket{x}\ket{0}\ket{q}\ket{0}\ket{q_t}\to \ket{x}(\otimes_i\ket{x_i}^{\otimes t_i})\ket{q}\ket{0}\ket{q_t},
\end{align}
where the number $t_i$ is determined by the input Boolean function. The depth of Copy stage is $\log_2(\max_i{t_i})=O(\log \ell)$.

In the Clause stage, we use the information in the $q_{mem}$ register and input qubits to synthesize clauses in parallel. The result is stored in the first half of $q_{clean}$ register.
\begin{align*}
    &\mathcal{CL}\ket{x}(\otimes_i\ket{x_i}^{\otimes t_i})\ket{q}\ket{0}\ket{q_t}
    \\\to& \ket{x}(\otimes_i\ket{x_i}^{\otimes t_i})\ket{q}\left(\otimes_{i=1}^{\lfloor\ell/2S\rfloor}\ket{C_{2i-1}\wedge C_{2i}}\right)\ket{q_t}.
\end{align*}
We can synthesize $O(\ell/k)$ terms with $O(k)$-depth circuit. The total depth of Clause stage is $O(k\log \ell)$.

In the merge stage, we merge all the clauses stored in the $q_{clean}$ to the target qubit.
\begin{align*}
    &\mathcal{ME}\left(\otimes_{i=1}^{\lfloor\ell/2S\rfloor}\ket{C_{2i-1}\wedge C_{2i}}\right)\ket{q_t}
    \\\to& \left(\otimes_{i=1}^{\lfloor\ell/2S\rfloor}\ket{C_{2i-1}\wedge C_{2i}}\right)\ket{q_t\oplus f(x)}.
\end{align*}
A Toffoli gate can merge two CNF formulae on a clean ancillary qubit. To merge $\ell/2(S+1)$ CNF formulae, the depth of merge stage is $O(\log(\ell/2(S+1)))=O(\log{l}).$

We repeat the Copy stage and the Clause stage to reset all the ancillary qubits in the reset stage.
\begin{align*}
    &\mathcal{RE}\ket{x}(\otimes_i\ket{x_i}^{\otimes t_i})\ket{q}\left(\otimes_{i=1}^{\lfloor\ell/2S\rfloor}\ket{C_{2i-1}\wedge C_{2i}}\right)\ket{q_t\oplus f(x)}\\
    \to& \ket{x}\ket{0}\ket{q}\ket{0}\ket{q_t\oplus f(x)}.
\end{align*}
We repeat the first two stages, and the depth of the reset stage is $O(k\log \ell)$.

The depth of inner-most recursion circuit is $O(k\log \ell)$. In the inner-most recursion, $\ell/S$ clauses can be synthesized in parallel. We use $\text{Depth}'_{\ell}(f)$ to denote the depth of the circuit, obtained by the algorithm described in this section, to synthesize $f$ with $\ell$ ancillary qubits. $\text{Depth}'_{\ell}(\CNF_{n,\ell/S}^{k})=O(k\log \ell)$. Combine with the recurrence formula in the previous section: $\text{Depth}'_{\ell}(\CNF_{n,m}^{k})=\frac{2\ell}{(S+1)}\text{Depth}'_{\ell}(\CNF_{n,2m(S+1)/\ell}^{k}).$ Then we have:
\[\text{Depth}'_{\ell}(\CNF_{n,m}^{k})=O\left(k\log\ell \left(\frac{mS}{\ell}\right)^{1+\log_{\ell/S} 4}\right).\]

\begin{figure*}[ht]
    \centerline{
    \Qcircuit @C=0.5em @R=.5em{
    \lstick{\ket{x}}&\qw /&\qw &\multigate{2}{\mathcal{G}_{2i-1}^{p_{2i-1}}}&\qw &\multigate{3}{\mathcal{G}_{2i}^{p_{2i}}}&\qw &\multigate{2}{\mathcal{G}_{2i-1}^{p_{2i-1}}}&\qw &\multigate{3}{\mathcal{G}_{2i}^{p_{2i}}}&\qw&\rstick{\ket{x}}\\
    \lstick{q_{mem},\ket{x'}}&\qw /&\qw &\ghost{\mathcal{G}_{2i-1}^{p_{2i-1}}}&\qw &\ghost{\mathcal{G}_{2i}^{p_{2i}}}&\qw &\ghost{\mathcal{G}_{2i-1}^{p_{2i-1}}}&\qw &\ghost{\mathcal{G}_{2i}^{p_{2i}}}&\qw&\rstick{\ket{x'}}\\
    \lstick{\ket{p_{2i-1}}}&\qw&\ctrl{1} &\ghost{\mathcal{G}_{2i-1}^{p_{2i-1}}}&\ctrl{1} &\ghost{\mathcal{G}_{2i}^{p_{2i}}}&\ctrl{1} &\ghost{\mathcal{G}_{2i-1}^{p_{2i-1}}}&\ctrl{1} &\ghost{\mathcal{G}_{2i}^{p_{2i}}}&\qw&\rstick{\ket{p_{2i-1}}}\\
    \lstick{\ket{p_{2i}}}&\qw &\ctrl{1} &\qw&\ctrl{1} &\ghost{\mathcal{G}_{2i}^{p_{2i}}}&\ctrl{1} &\qw&\ctrl{1} &\ghost{\mathcal{G}_{2i}^{p_{2i}}}&\qw&\rstick{\ket{p_{2i}}}\\
    \lstick{\ket{q_{i}}}&\qw&\targ &\qw&\targ &\qw&\targ &\qw&\targ &\qw&\qw&\rstick{\ket{q_i\oplus(C_{2i-1}\vee C_{2i})}}\\
    }}
	\caption{The Clause stage in the inner-most recursion of algorithm. The $q_{dirty}$ and $q_{clean}$ registers are divided into $\ell/2S$ parts. Here we show the $i$-th part of Clause stage, where we synthesize a two clauses function. Different part of Clause is running in parallel. Operator $\mathcal{G}_j^{q_t}$ generate a clause $C_j$ on the target qubit $q_t$, $\mathcal{G}_j^{q_t}\ket{x}\ket{q_t}\to \ket{x}\ket{q_t\oplus C_j(x)}.$}
	\label{fig:inner_clause}
\end{figure*}
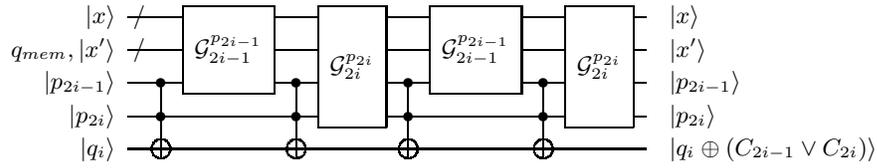
\subsection{Circuit lower bound for CNF synthesis}
We will prove that there exists a  $k$-CNF with $m$ clauses which need $\Omega(km)$ size of quantum circuits to approximate it with any error $\varepsilon < \frac{\sqrt{2}}{2}$, as depicted in Theorem \ref{thm:sizeLowerBound}.
To obtain this lower bound, let us first give a lower bound for the number of different ${\rm CNF}_{n,m}^k$.

\begin{lemma}
There are $\Omega\pbra{\binom{\binom{n}{k}}{m}}$ different instances for ${\rm CNF}_{n,m}^k$.
\label{lem:kCNFLower}
\end{lemma}
\begin{proof} 
Note that a CNF formula $\phi: \{T,F\}^n \rightarrow \{T,F\}$ can be uniquely represented as a Boolean function $f_{\phi}:\{0,1\}^n\rightarrow \{0,1\}$. With a little abuse of symbols, we use the same symbol to represent the input of CNF formula and the corresponding Boolean functions.
Let 
$\phi = \pbra{v_{1}\vee \cdots \vee v_{k}} 
\wedge \dots \wedge \pbra{v_{(k-1)m+1}\vee \cdots \vee v_{km}}$
be a $k$-CNF formula, where $v_i\in\{x_1,\ldots, x_n,\neg x_1, \ldots, \neg x_n\}$, then it can be represented as a Boolean function 
\begin{equation}
    f_{\phi}(x) = \pbra{1-\prod_{j = 1}^k  \bar{v}_{j}}
    \cdots \pbra{1-\prod_{j = 1}^{k} \bar{v}_{(k-1)m +j}},
    \label{eq:boolRep}
\end{equation}
where 
$\bar{v}_{i} = 0$ iff $v_{i} = T$ and $\bar{v_{i}} = 1$ 
iff $v_{i} = F$, and the input $x_i\in\{0,1\}$ of $f_{\phi}$ is associated with $x_i\in\{T,F\}$ of $\phi$.
Let
$f(x),g(x)$ be two functions (formulas), we say $f \equiv g$ if $f(x_1,\cdots, x_n) = g(x_1,\cdots, x_n)$ for any legal input $(x_1,\cdots, x_n)$.
Let 
\[\Lcal := \cbra{v_1 \vee \cdots\vee v_k| v_i\in \{ \neg x_1,\ldots, \neg{x_n}\}, v_i \ne v_j\text{ for } i,j\in[k]}\]
be the set of all clauses with $k$-variables, where ${x}_j\in\{T,F\}$, and $\neg x_j$ are the negations of $x_j$.
Let a set of $k$-CNF formulas be
\[\Acal := \{\phi| \phi = l_1 \wedge \cdots \wedge l_m, l_i \in \Lcal, l_i \ne l_j, \text{ for }i\ne j\in[m]\}.\]
We would like to show the size of $\Acal$ is size$(\Acal) = \binom{\binom{n}{k}}{m}$. \emph{i.e.}, any two different formulas $\phi_{1}=l_1\wedge \ldots \wedge l_m, \phi_{2}=l_1'\wedge \ldots \wedge l'_m$ such that there exists $j\in[m]$, $l_j\ne l_j'$, we have $\phi_{1}\not\equiv \phi_{2}$.
 By contradiction, suppose $\phi_{1} \equiv \phi_{2}$, then $f_{\phi_{1}}(x) \equiv f_{\phi_{2}}(x)$. 
 Let $g_{\phi}(x)$ satisfies deg$(g)=k$ be the summations of all degree-$k$ terms of $f_{\phi}(x)$.
By the definition of $f_{\phi}(x)$ in Equation \eqref{eq:boolRep} and the fact that $\bar{v}_j^2 = \bar{v}_j$, 
\[g_{\phi} (x) = -\sum_{j=0}^{m-1}\bar{v}_{jk+1} \cdots \bar v_{j(k+1)}.\]
Since $f_{\phi_{1}}(x) \equiv f_{\phi_{2}}(x)$, then deg$(f_{\phi_{1}} - f_{\phi_{2}}) = 0$, \emph{i.e.}, $g_{\phi_{1}}(x) \equiv g_{\phi_{2}}(x)$.
 Let $ \bar{v}_1  \cdots \bar{v}_k$ be one term of $g_{\phi_1}(x)$. For a given input $x = \pbra{x_1,\ldots, x_n}$ such that $x_i = 1$ when $x_i\in \{\bar{v}_1, \dots, \bar{v}_k\}$ and $x_i = 0$ otherwise. It is easy to check
 $g_{\phi_2}(x) = 1$ iff $x_1\cdots x_k$ is one term of the function $g_\phi(x)$. Hence $\bar{v}_1  \cdots \bar{v}_k$ is also a term of $g_{\phi_2}(x)$.
Without loss of generality, each term $\bar{l}_{j}= \bar{v}_{jk+1} \cdots \bar v_{j(k+1)}\in g_{\phi_{1}}$, $\bar{l}_{j} \in g_{\phi_{2}}$ at the same time. Therefore $\phi_{1} = \phi_{2}$, contrary with the fact that $\phi_{1},\phi_{2}$ are two different formulas in $\Acal$.

\end{proof}



\begin{theorem}
There exists a ${\rm CNF}_{n,m}^k$, any quantum circuits approximating it with error $\varepsilon < \frac{\sqrt{2}}{2}$ needs size $\Omega\pbra{km}$.
\label{thm:sizeLowerBound}
\end{theorem}
\begin{proof} 
 Let $U\in\Cbb^{4\times 4}$ be a two qubit gate, and the $\delta$-discretization of the $(j,k)$-th element $U_{jk}$ be $U_{jk}^{\delta} = \delta \lfloor a/\delta\rfloor + i\delta \lfloor b/\delta\rfloor$, where $U_{jk} = a+ib$. Then we have $\vabs{U -U ^{\delta}}_2< 2\delta$. There are at most $ \pbra{\frac{2}{\delta}}^{32}$ different $\delta$-discretizations $U ^{\delta}$ for the infinite continuous $U $ in the space by its definition. 

 In the following, we prove that any two different instances in $\Acal$ do not share any common $\delta$-discretization. Hence, we can use the counting method to give a lower bound of the circuit size.
 
 Let $A_{G}, A_{H}$ be the quantum circuit representations of two different instances in $\Acal$. Let $s$ be the maximum size of all the unitaries related to $A_{G}$ and $A_{H}$. 
 By the fact that the unitary $U \in \Cbb^{2^{n}\times 2^{n}}$ has a $s$-size quantum circuit,
 the following inequalities
 \begin{align*}
 \begin{aligned}
 &\vabs{A_{G}- A_{G}^{\delta}} < 2s\delta\leq \varepsilon,\\
 &\vabs{A_{H}- A_{H}^{\delta}} < 2s\delta\leq \varepsilon,\\
 \end{aligned}
 \end{align*}
 hold when $\delta = \frac{ \varepsilon}{2s}$, and $\varepsilon<\frac{\sqrt{2}}{2}$.
 Combined with the fact that $\vabs{A_{G} - A_{H}} = \sqrt{2}$, we have $A_{G}^{\delta}\ne A_{H}^{\delta}$.

  Hence, any two different instances in $\Acal$ have different $\delta$-discretization.
 There are $\Omega\pbra{\binom{\binom{n}{k}}{m}}$ different instances for $k$-CNF with $m$-clause by Lemma 3 in main file.
 By the fact that the number of different instances is upper bounded by the number of $\delta$-discretization of quantum circuits,
  \begin{equation}
      \binom{\binom{n}{k}}{m} \leq \pbra{\pbra{\frac{2}{\delta}}^{32}\cdot n}^s.
      \label{eq:CountDis}
  \end{equation} 
  Since
  $\binom{n}{k} = \Omega ((n/k)^k)$ for any $k$. Then
  \begin{align*}
      \begin{aligned}
      \binom{\binom{n}{k}}{m} = \Omega\pbra{\pbra{\frac{\binom{n}{k}}{m}}^{m} } = \Omega\pbra{\pbra{\frac{n^{k}}{k^km}}^{m}}.
      \end{aligned}
  \end{align*}
  By inequality \ref{eq:CountDis}, we have $s = \Omega(km)$ when $k=o(n)$.
  When $k = cn$ for constant $c<1$, by Stirling's formula, $\binom{n}{k} = \Omega\pbra{2^{an}}$ for some constant $a<1$. Hence,
\begin{align*}
    \binom{\binom{n}{k}}{m} = \Omega\pbra{\pbra{\frac{2^{an}}{m}}^m},
\end{align*}
combined with inequality \eqref{eq:CountDis} give us $s = \Omega(km)$ when $k = cn$ for constant $c<1$. This implies the lower bound also holds for any $k<n$ for general $k$-CNF.
\end{proof}

\bibliographystyle{plainnat}
\bibliography{ref}